\documentstyle[prl,aps,floats,epsfig,twocolumn]{revtex}
\tighten
\let\jnfont=\rm
\def\NPB#1,{{\jnfont Nucl.\ Phys.\ }{\bf B#1},}
\def\PLB#1,{{\jnfont Phys.\ Lett.\ B }{\bf #1},}
\def\PRD#1,{{\jnfont Phys.\ Rev.\ D }{\bf #1},}
\def\PRL#1,{{\jnfont Phys.\ Rev.\ Lett.\ }{\bf #1},}
\def\ZPC#1,{{\jnfont Z.~Phys.\ C }{\bf #1},}
\begin{document}
\draft
\preprint{}

\title{Split Two-Higgs-Doublet Model and Neutrino Condensation}

\author{Fei Wang $^1$, Wenyu Wang $^1$, Jin Min Yang $^{2,1}$ }

\address{ \ \\[2mm]
{\it $^1$ Institute of Theoretical Physics, Academia Sinica, Beijing 100080, China} \\ [2mm]
{\it $^2$ CCAST(World Laboratory), P.O.Box 8730, Beijing 100080, China} \ \\[6mm] }

\maketitle

\begin{abstract}
We split the two-Higgs-doublet model by assuming very different vevs for the
two doublets: the vev is at weak scale (174 GeV) for the doublet $\Phi_1$
and at neutrino-mass scale ($ 10^{-2} \sim10^{-3}$ eV) for the doublet $\Phi_2$.
$\Phi_1$ is responsible for giving masses to all fermions except
neutrinos; while $\Phi_2$ is responsible for giving neutrino masses
through its tiny vev without introducing see-saw mechanism.
Among the predicted five physical scalars $H$, $h$,  $A^0$ and $H^{\pm}$,
the CP-even scalar $h$ is as light as  $10^{-2} \sim 10^{-3}$ eV while others are
at weak scale.
We identify $h$ as the cosmic dark energy field and the other CP-even scalar
$H$ as the Standard Model Higgs boson;  while the CP-odd $A^0$ and
the charged $H^{\pm}$ are the exotic scalars to be discovered
at future colliders. Also we demonstrate a possible dynamical origin for the doublet
$\Phi_2$ from neutrino condensation caused by some unknown dynamics.
\end{abstract}


{\em Introduction:~~} The two-Higgs-doublet model (2HDM)
\cite{2hdm} is the most naive extension of the Standard Model
(SM). However, such an extension seems to bring more heat than
light: it plainly introduces several more scalar particles without
solving any problems in the SM. In this work we reform this model
in order to explain the tiny neutrino masses without see-saw
mechanism and provide a candidate field for cosmic dark energy.
For this purpose we split this model by  assuming very different
vevs for the two doublets: the vev is at weak scale (174 GeV) for
the doublet $\Phi_1$ and at neutrino-mass scale  ($ 10^{-2}
\sim10^{-3}$ eV) for the doublet $\Phi_2$. We assume that $\Phi_1$
is responsible for giving masses to all fermions except neutrinos;
while $\Phi_2$ is responsible for giving neutrino masses through
its tiny vev without introducing see-saw mechanism. Among the
predicted five physical scalars $H$, $h$,  $A^0$ and $H^{\pm}$,
the CP-even scalar $h$ is as light as  $10^{-2} \sim 10^{-3}$ eV
(in the following we assume $10^{-3}$ eV for example) while others 
are at weak scale. We identify $h$ as the cosmic dark
energy field and the other CP-even $H$ as the SM Higgs boson;
while the CP-odd $A^0$ and the charged $H^{\pm}$ are the exotic
scalars to be discovered at future colliders.

Our motivations for splitting the 2HDM are then quite clear:
\begin{itemize}
\item  By identifying $h$ as the cosmic dark energy field, we try to give an explanation
for dark energy (although we know we may be quite far from the true story).
As is well known, the dark energy seems to be a great mystery in today's physics and cosmology.
The precise cosmological measurements, such as the Wilkinson Microwave Anisotropy
Probe (WMAP) measurements \cite{wmap},  indicate that the major content of today's universe
is the weird dark energy. So far we lack the understanding about the nature of dark energy,
although some phenomenological approaches have been proposed \cite{quint}.

\item  By giving neutrino masses through the tiny vev of $\Phi_2$, we try to {\it 'understand'}
the tiny neutrino masses without introducing see-saw mechanism (although we know the
see-saw mechanism is quite elegant).
It is also well known that among the elementary particles the neutrinos are quite
special species due to their extremely small masses.
Without see-saw mechanism, the Yukawa couplings of neutrinos
with the SM Higgs doublet are extremely small, which is hard to understand.
We may speculate that the origin of
neutrino masses are different from other fermions, i.e., the neutrino masses are not from
the Yukawa couplings of the SM Higgs doublet (there might be some symmetry to
forbid the neutrino couplings with the SM Higgs doublet), and, instead, they are from the
Yukawa couplings of a new scalar doublet which has a tiny vev.

\item We try to relate neutrino mass generation with dark energy puzzle.
The neutrino mass scale is seemingly near or coincident with the cosmic dark energy
scale (the dark energy density is $\sim (10^{-3} eV)^4$ and thus the dark energy scale is
$\sim 10^{-3} eV$). Such an seeming coincidence has already stimulated some speculations
on the possible relation between neutrino and dark energy \cite{mass-varying}.
\end{itemize}

While the dynamical origin of the scalar doublet $\Phi_1$ with a vev at weak scale may be
something like technicolor, we propose that the scalar doublet $\Phi_2$ with a tiny vev at
the neutrino-mass scale may be from neutrino condensation.
Similar to the idea of top-quark condensation \cite{condensation}, we assume that
a four-fermion interaction for the third-family neutrino is induced at some high energy scale
(say TeV) from some unknown new dynamics (like top color \cite{top-color}) which is strong enough
to cause neutrino condensation
\footnote{Note that in the literature the neutrino condensation was once proposed as a
dynamical electroweak symmetry breaking mechanism \cite{lindner}, which can generate tiny
neutrino mass by incorporating the see-saw mechanism.}.

{\em Split two-Higgs-doublet model:~~}
We introduce two scalar
doublets $\Phi_1$ and $\Phi_2$ as
\begin{eqnarray} \label{vev1}
\Phi_1&=&\left( \begin{array}{c}\phi_1^+ \\ {\rm Re}\phi_1^0+v_1+i {\rm Im} \phi_1^0
              \end{array} \right) , \\ \label{vev2}
\Phi_2&=&\left( \begin{array}{c}\phi_2^+ \\ {\rm Re}\phi_2^0+v_2+i {\rm Im} \phi_2^0
              \end{array} \right) .
\end{eqnarray}
We split the two vevs as
\begin{eqnarray}
 v_1 \sim 174~ {\rm GeV}, ~~~~
 v_2 \sim  10^{-3}~ {\rm eV} .
\end{eqnarray}
By assumption,  $\Phi_1$ is responsible for giving masses to all fermions except
neutrinos;  while $\Phi_2$ is responsible for giving neutrino masses through its
tiny vev. Of course, both of them contribute to $W$ boson mass:
\begin{eqnarray}
m_W^2=g^2(v_1^2+v_2^2)/2 \approx g^2 v_1^2 /2 .
\end{eqnarray}
We assume CP conservation and discrete symmetry $\Phi_1 \to -\Phi_1$ for the
potential of the scalars. Then the general potential takes the form \cite{2hdm}
\begin{eqnarray} \label{mix}
V(\Phi_1,\Phi_2)&=& \lambda_1(\Phi_1^\dagger \Phi_1-v_1^2)^2
                     +\lambda_2(\Phi_2^\dagger \Phi_2-v_2^2)^2 \nonumber \\
  && +\lambda_3 \left[(\Phi_1^\dagger \Phi_1-v_1^2)+(\Phi_2^\dagger \Phi_2-v_2^2) \right]^2
     \nonumber \\
  && +\lambda_4 \left[(\Phi_1^\dagger \Phi_1)(\Phi_2^\dagger \Phi_2)
                     -(\Phi_1^\dagger \Phi_2)(\Phi_2^\dagger \Phi_1)\right]
     \nonumber \\
  &&
     + \lambda_6 \left[{\rm Im} (\Phi_1^\dagger \Phi_2) \right]^2,
\end{eqnarray}
where $\lambda_i$ are non-negative real parameters.
Note that since we assumed the exact discrete symmetry $\Phi_1 \to -\Phi_1$, we
dropped the term
\begin{eqnarray} \label{term5}
\lambda_5 \left[{\rm Re}~(\Phi_1^\dagger \Phi_2)-v_1 v_2 \right]^2 ,
\end{eqnarray}
which softly breaks such a discrete symmetry.

It should be pointed out that since the two vevs of the potential in eq.(\ref{mix})
are splitted by many orders, in order to make it stable under radiative 
corrections we either need an extreme fine tuning
as that occuring in GUTs or introduce some fancy symmerty to stablize it.    

After the diagonalization of the mass-square matrices in Eq.(\ref{mix}),
we obtain eight mass eigenstates: $H$, $h$, $H^{\pm}$, $G^{\pm}$, $A^0$ and $G^0$,
among which  $H$ is identified as the SM physical Higgs boson, and
$G^{\pm}$ and $G^0$ are massless Goldstone boson eaten by $W$ and $Z$ gauge bosons.

The two charged scalars $H^{\pm}$ and $G^{\pm}$ are obtained by
\begin{eqnarray}
G^{\pm} &=&  \Phi_1^{\pm} \cos\beta + \Phi_2^{\pm} \sin\beta \approx  \Phi_1^{\pm} ,\\
H^{\pm} &=& -\Phi_1^{\pm} \sin\beta + \Phi_2^{\pm} \cos\beta  \approx \Phi_2^{\pm} ,
\end{eqnarray}
while the two CP-odd neutral scalars $A^0$ and $G^0$ are obtained by
\begin{eqnarray}
G^0 &=&  \sqrt{2}({\rm Im} \Phi_1^0 \cos\beta +{\rm Im} \Phi_2^0 \sin\beta)
        \approx  \sqrt{2}{\rm Im} \Phi_1^0 ,\\
A^0 &=&  \sqrt{2}(-{\rm Im} \Phi_1^0 \sin\beta +{\rm Im} \Phi_2^0 \cos\beta)
        \approx  \sqrt{2}{\rm Im} \Phi_2^0 ,
\end{eqnarray}
where the mixing angle $\beta$ is very small since $\tan\beta = v_2/v_1$.

The masses of $H^{\pm}$ and  $A^0$ are given by
\begin{eqnarray} \label{ch-mass}
m_{H^{\pm}}^2 &=& \lambda_4 (v_1^2+v_2^2) \approx  \lambda_4 v_1^2 ,\\
m_{A^0}^2     &=& \lambda_6 (v_1^2+v_2^2) \approx  \lambda_6 v_1^2 .
\end{eqnarray}
So, both $m_{H^{\pm}}$ and $m_{A^0}$ should be at weak scale given that
$\lambda_i\sim {\cal O}(1)$.

The two CP-even scalars $H$ and $h$ are obtained by
\begin{eqnarray}
H &=&  \sqrt{2} \left[ ({\rm Re}~\Phi_1^0-v_1) \cos\alpha + ({\rm Re}~\Phi_2^0-v_2) \sin\alpha
                \right] \nonumber \\
                &\approx&  \sqrt{2}  ({\rm Re}~\Phi_1^0-v_1) , \\
h &=&  \sqrt{2} \left[-({\rm Re}~\Phi_1^0-v_1) \sin\alpha + ({\rm Re}~\Phi_2^0-v_2) \cos\alpha
                \right]\nonumber \\
                &\approx& \sqrt{2}  ({\rm Re}~\Phi_2^0-v_2) .
\end{eqnarray}
Here the mixing angle $\alpha$ is also very small since
\begin{eqnarray}
\tan(2\alpha) = \frac{2 {\cal M}_{12}}{ {\cal M}_{11}- {\cal M}_{22}} \sim v_2/v_1 ,
\end{eqnarray}
where $ {\cal M}_{ij}$ are elements of the mass-square matrix
\begin{eqnarray}
 {\cal M}=\left( \begin{array}{cc} 4(\lambda_1+\lambda_3) v_1^2 & 4 \lambda_3 v_1 v_2 \\
                                   4 \lambda_3 v_1 v_2 & 4(\lambda_2+\lambda_3) v_2^2
                 \end{array} \right) .
\end{eqnarray}
The masses of $H$ and $h$ are given by
\begin{eqnarray}
m_H^2 &\approx &{\cal M}_{11}= 4(\lambda_1+\lambda_3) v_1^2, \\
\label{mxh}
m_h^2 &\approx &{\cal M}_{22}-\frac{{\cal M}_{12}^2}{{\cal M}_{11}}
      =4\left[(\lambda_2+\lambda_3)-\frac{\lambda_3^2}{(\lambda_1+\lambda_3)}\right] v_2^2 .
\end{eqnarray}

Therefore, among the five physical scalars,  the masses of $H^{\pm}$, $A^0$ and $H$ are at
the weak scale since they are proportional to $v_1\approx 174$ GeV, while the mass of  $h$
is at the neutrino mass scale since it is proportional to $v_2\sim 10^{-3}$ eV. The two
mixing angles  $\alpha$ and $\beta$ are very small since they are  proportional to $v_2/v_1$.

Due to the negligibly small mixing angles  $\alpha$ and $\beta$, the
properties of these scalars are approximately like this:
\begin{itemize}
\item[(i)]   For $H$: All its couplings are the same as in the SM.
\item[(ii)]  For  $H^{\pm}$: They have Yukawa couplings only to
             $\bar\ell_L \nu^\ell_R$ ($\ell=e,\mu,\tau$), but
             the coupling strength is at a natural order, say ${\cal O}(1)$.
             Their gauge couplings like $H^+H^-\gamma$ and  $H^+H^-Z$ are
             the same as in the usual 2HDM.
\item[(iii)] For $A^0$: It has Yukawa couplings only to neutrino pairs.
             It has gauge couplings like $ZhA^0$ as in the usual 2HDM.
\item[(iv)]  For $h$:  It has Yukawa couplings only to neutrino pairs.
             Its gauge couplings to $W^+W^-$ and  $ZZ$ are very weak
             (proportional to $v_2$).
\end{itemize}
Therefore, the experimental constraints on these scalars are:
\begin{itemize}
\item[(1)]  For $H$:  all direct and indirect experimental constraints are the same as
            for the SM Higgs boson.
\item[(2)]  For  $H^{\pm}$ and  $A^0$:  the experimental constraints are
            similar as in the usual 2HDM except for the invalid constraints
            from various $B$-decays (such as $b\to s \gamma$).
            For example, from the unobservation of $e^+e^- \to H^+H^-$ at LEP II,
            $H^{\pm}$ should be heavier than about 100 GeV and thus $\lambda_4$
            in Eq. (\ref{ch-mass}) cannot be too small.
\item[(3)]  For the ultra-light $h$: stringent constraints both
            from particle physics and from astrophysics are derived from
            its interactions with either photons, electrons or nucleons \cite{pdg}.
            For example, stringent constraints may come from positronium decays,
            meson decays, quarkonium decays or nuclear transitions.
            In our model, fortunately, these constraints can be avoided or become
            quite weak since the coupling of $h$ with photons, electrons or nucleons
            are suppressed by $v_2/v_1\sim 10^{-14}$ at tree-level. The most dangerous
            constraints may come from the invisible $Z$ decays. For example, from the three-body
            decay $Z\to h (A^0)^* \to h \nu \bar\nu$, some lower mass bound (say TeV) may be
            set on $A^0$.

          Note that the light scalar $h$ has a Yukawa coupling of order one to neutrinos
          (left and right handed). Thus, in addition to left-handed neutrinos, the right-handed
          neutrinos and  the scalar $h$ also appear in the thermal equilibrium in the early
          universe. This means that the effective total number of neutrino species ($N_\nu$)
          is about 6.5 instead of the standard number of 3. We should check if this is allowed by
          cosmology and astrophysics (right-handed neutrinos are subject to no constraints from
          particle physics experiments such as LEP experiments since they are gauge singlet
          and have no gauge couplings).
          Firstly, we note that stringent constraint on $N_\nu$
          has been derived from the standard BBN \cite{BBN}.
          However, as discussed in \cite{DBBN}, such
          BBN contstraint can be  relaxed  since it is obtained under the assumption that
          the chemical potential for the background neutrinos is negligible. The relaxed $2\sigma$
          bound from the combined analysis of BBN, CMB and supernova data is $N_\nu < 7$
          \cite{DBBN}. Secondly, we should seriously and specially consider the constraints
          from SN1987A since the additional neutrino species can speed up the cooling-down
          of its core. It is well known that there exist some uncertainty for the estimation
          of the total binding energy as well as the effective temperature of the supernova.
          As analysed in \cite{ellis}, the constraint is $N_\nu < 6.7$ (the Eq.8
          in \cite{ellis}).

          It should also be noted that since the light scalar $h$ can couple to ordinary matter
          through the mixing with the SM Higgs boson $H$ (suppressed by $v_2/v_1\sim 10^{-14}$)
          or radiatively through loops involving left- and right-handed neutrinos plus weak 
          gauge bosons (also suppressed by $v_2/v_1$ because right-handed neutrinos do not
          have any gauge interactions and a neutrino mass insertion is needed in order to 
          couple to gauge bosons), it is necessary to check whether the gravitational 
          equivalence principle is still respected in our case. 
          As analysed in \cite{h-bound}, it will be consistent with 
          tests of the gravitational inverse square law as well as the equivalence principle
          as long as the scalar is not too light or its 
          Yukawa couplings to ordinary matter are sufficiently weak. 
          Following the analysis in \cite{h-bound}, we checked that our case is marginally 
          in the region allowed by the gravitational inverse square law and the equivalence 
          principle.    
\end{itemize}

{\em Neutrino Condensation:~~} Now we try to provide an
explanation for the generation of $\Phi_2$ by neutrino
condensation. We assume that the third-family leptons have an
effective four-fermion interaction, which may be generated at some
high energy scale $\Lambda$ above TeV from some unknown new
dynamics. The underlying new dynamics, which is not specified
here, might be some non-abelian gauge interaction spontaneously
broken at some higher scale. So such new dynamics causes
negligible effects to the electroweak physics of the third-family
leptons.

The four-fermion effective interaction for the third-family leptons takes the form
at the energy scale $\Lambda$
\begin{equation}
  G \left( \bar\Psi_L \nu_R\right)\left( \bar{\nu}_R{\Psi}_L\right) ,
\end{equation}
where $G$ is the coupling constant, running with the energy scale.
$\Psi_L$ is the doublet of the left-handed third-family lepton fields and
$\nu_R$ is the right-handed tau-neutrino.
When $G \Lambda^2\gg 1$, the tau-neutrinos condensate
and the condensation effects can be incorporated by introducing
an auxiliary scalar field  $\Phi_2$ into the Lagrangian
\begin{eqnarray}
  -M_0 \sqrt{G}\left[ \bar\Psi_L \tilde\Phi_2 \nu_R +h.c. \right]
      -M_0^2 \tilde\Phi_2^{\dag}\tilde\Phi_2
\end{eqnarray}
where $M_0$ is an unspecified bare mass parameter and $\tilde\Phi_2= i {\tau}_2 \Phi_2^*$.
$\Phi_2$ and $\tilde\Phi_2$ take the form
\begin{eqnarray}
\Phi_2 &=& \frac{\sqrt{G}}{2 M_0}\left( \begin{array}{c}
                                         \bar{\tau}(1+\gamma_5)\nu \\
                                        -\bar{\nu}(1+\gamma_5)\nu
                                       \end{array}\right)
              =\frac{\sqrt{G}}{M_0}(i {\tau}_2) {(\bar{\nu}_R {\Psi}_L)}^{\dagger T} \\
\tilde\Phi_2 &=&  -\frac{\sqrt{G}}{M_0} ( \bar{\nu}_R {\Psi}_L) .
\end{eqnarray}
As the energy scale runs down, such an auxiliary field $\Phi_2$ gets gauge
invariant kinematic terms as well as  quartic interactions through quantum effects:
\begin{eqnarray}
&& -M_0 \sqrt{G} \left(\bar\Psi_L \tilde\Phi_2\nu_R +h.c. \right)
    +Z_{\Phi_2}|D_{\mu} \Phi_2|^2 \nonumber\\
&&  -M_{\Phi_2}^2 \Phi_2^\dag  \Phi_2-\lambda_0 \left(\Phi_2^\dag
\Phi_2\right)^2 ,
\end{eqnarray}
where
\begin{eqnarray}
 Z_{\Phi_2}  &=& \frac{N_c M_0^2 G}{{(4 \pi)}^2 }
                            \ln\left(\frac{{\Lambda}^2}{{\mu}^2}\right)\\
 M_{\Phi_2}^2&=& M_0^2-\frac{2 N_c M_0^2 G}{{(4 \pi)}^2 }({\Lambda}^2-{\mu}^2) \\
 \lambda_0 &=& \frac{N_c {(M_0^2 G)}^2 }{{(4 \pi)}^2 }
                            \ln\left( \frac{{\Lambda}^2}{{\mu}^2}\right) .
\end{eqnarray}
Here $N_C$ is the 'color' number of tau-neutrino in the unspecified new dynamics.
Redefining $\sqrt{Z_{\phi_2}}\Phi_2$ as $\Phi_2$ and $g_t= M_0 \sqrt{G}/\sqrt{Z_{\Phi_2}}$,
$m^2= M_{\Phi_2}^2/ Z_{\Phi_2}$, $\lambda=\lambda_0/Z_{\Phi_2}^2$, we obtain the
Lagrangian
\begin{eqnarray} \label{potential}
&&  -g_t\left(\bar{\Psi}_L \tilde\Phi_2\nu_R +h.c. \right)
  +|D_{\mu} \Phi_2|^2 \nonumber\\
&& -m^2 \left( \Phi_2^\dag \Phi_2 \right)-\lambda
{\left(\Phi_2^\dag \Phi_2 \right) }^2
\end{eqnarray}
For $\mu\ll\Lambda$, we have $m^2<0$ when
$G \geq G_{crit}=8 {\pi}^2/(N_c {\Lambda}^2)$, which leads to spontaneous breaking of
electroweak symmetry with $\Phi_2$ developing a vev $v_2=\sqrt{-m^2/2\lambda}$.

The masses of tau-neutrino is given by
\begin{eqnarray}
\label{mass}
m_{\nu}^2=g_t^2v_2^2
         =\frac{16 {\pi}^2 v_2^2}{ N_c \ln({\Lambda}^2/ m_{\nu}^2 ) } ,
\end{eqnarray}
which is consistent with $m_\nu\sim v_2 \sim 10^{-3}$ eV for
$\Lambda \sim$ TeV.

{\em Collider and Cosmological Consequences:~~} First, we briefly
discuss the phenomenology at the colliders, LHC (CERN Large Hadron
Collider) and ILC (International Linear Collider). Since the
phenomenology of $H$ is approximately the same as the SM Higgs
boson, we focus on the phenomenology of $H^{\pm}$, $A^0$ and $h$.
\begin{itemize}
\item[(1)] Since $h$ is as light as neutrino and couples with neutrinos, its
           dominant decay mode is
           $h\to \nu\bar\nu$. Thus it just escapes detection at colliders although
           it can be produced through $e^+e^-\to Z^* \to h A^0$ at the ILC.
\item[(2)] At the ILC $A^0$ can be produced through $e^+e^-\to Z^* \to h A^0$ followed by
           the decay  $A^0\to Z h$, leading to the signature of two lepton or
           two jets plus missing energy.
\item[(3)] At the ILC $H^{\pm}$ can be pair produced not only through the usual s-channel
           process $e^+e^-\to$ $Z^*$, $\gamma^*$ $\to$ $H^+ H^-$ but also through the t-channel
           process via exchanging a $\nu_R$. Due to such an additional  t-channel process,
           the production rate will be different from the usual 2HDM prediction. The
           decay modes  $H^-\to \ell_L \bar \nu_R$ ($\ell=e$,$\mu$) will give good signatures.
           Since $H^{\pm}$ almost do not couple to quarks, they cannot be produced
           at LHC collider through the subprocess like $gb\to t H^-$.
\end{itemize}

Now we turn to the cosmological consequences.
Since the CP-even physical scalar $h$ is an ultra-light scalar particle,
at the order of $10^{-3}$ eV, we may interpret this scalar field as  the dark energy field.
Since the corresponding vev $v_2$ is at the order of $10^{-3}$ eV, the magnitude of the
vacuum energy (dark energy) should be naturally around the order of ${(10^{-3}~eV)}^4$.
This means that in fact we assume something that sends to zero the
potential energy related to the usual Weinberg-Salam Higgs doublet 
while leaving the potential energy related to the light scalar $h$ unbalanced.

Since $h$ may decay into $\nu \bar\nu$, we wonder if it is still present
in today's universe as a viable dark energy field.
As is well known, there is now a neutrino microwave background at 1.9 K which corresponds to
$\sim 10^{-4}$ eV. The scalar $h$ and the neutrinos can reach thermal equilibrium if the
mass difference between $h$ its decay products $\nu \bar\nu$ is at the same order as the
neutrino background temperature.
Assuming  the contribution of chemical potential is at the order of unity, the ratio for
$n_h$ and $n_{\nu}$ in equilibrium is proportional to $exp(-\Delta Q/ T )$, where
$\Delta Q$ is the released energy in the decay process.

Note that just like some authors in \cite{mass-varying}, we may assume the dark fluid to be
the sum of the scalar potential of $h$ and the energy density in neutrino masses.  
In this case, since the scalar potential evolves with energy scale (time),  
the minimum of the resulting potential would evolve in time and also one 
could have mass varying neutrinos.

\end{document}